\newcommand{\ket}[1]{\ensuremath{\left| {#1} \right>}}
\documentclass[12pt]{iopart}
\usepackage[pdftex]{color,graphicx}
\usepackage{float}
\begin{document}
\title{Coherently displaced oscillator quantum states of a single trapped atom 
}
\author{Katherine C. McCormick$^{1,2}$, Jonas Keller$^{1,2}$, 
David J. Wineland$^{1,2,3}$, Andrew C. Wilson$^1$, Dietrich Leibfried$^1$ }

\address{$^1$National Institute of Standards and Technology, Boulder, CO 80305, USA}
\address{$^2$University of Colorado, Department of Physics, Boulder, CO 80305, USA}
\address{$^3$University of Oregon, Department of Physics, Eugene, OR 97403, USA}
\date{}
\begin{abstract}
Coherently displaced harmonic oscillator number states of a harmonically bound ion can be coupled to two internal states of the ion by a laser-induced motional sideband interaction. The internal states can subsequently be read out in a projective measurement via state-dependent fluorescence, with near-unit fidelity. This leads to a rich set of line shapes when recording the internal-state excitation probability after a sideband excitation, as a function of the frequency detuning of the displacement drive with respect to the ion's motional frequency. We precisely characterize the coherent displacement based on the resulting line shapes, which exhibit sharp features that are useful for oscillator frequency determination from the single quantum regime up to very large coherent states with average occupation numbers of several hundred. We also introduce a technique based on multiple coherent displacements and free precession for characterizing noise on the trapping potential in the frequency range of 500 Hz to 400 kHz. Signals from the ion  are directly used to find and eliminate sources of technical noise in this typically unaccessed part of the spectrum.
%
\end{abstract}
\maketitle
\section{Introduction}
As this focus issue on Quantum Optomechanics showcases, the manipulation of quantum states of a harmonic oscillator (HO) is a theme of current interest across a wide range of experimental platforms. Often the methods developed on one platform, in our case a single, harmonically bound atomic ion, can be adapted to many other platforms, after suitable modifications of the original procedures. An important example is resolved sideband-cooling of micro-fabricated oscillators, theoretically described in \cite{Wils07,Marq07}, that bears strong analogies to the methods that were first developed for single-ion mechanical oscillator systems \cite{Neuh78,Wine79,Died89}. In this contribution to the focus issue, ground-state cooling is combined with another basic idea that is widely applicable across HO platforms, namely coherent displacements that can be conveniently implemented with a classical force that is near resonant with the HO \cite{Carr65}. The final ingredient here is a suitable two-level system, in our case two internal electronic states of a single ion, that can be coupled to the HO motion and read out with a projective measurement to gain information about the state of the HO. A superconducting qubit is just one example of an effective two-level system that has been coupled to a co-located micro-fabricated HO \cite{Chu18}. Therefore, the methods described in this contribution might also be adaptable and useful to the rapidly growing community that studies HO systems in the quantum regime.\\
\\
Exciting the HO motion of trapped charged particles with a weak oscillating electric field, a method often called a ``tickle,'' has long been used to determine motional frequencies and subsequently the charge-to-mass ratio in various ion trap based devices, for example in ion-trap mass spectrometers \cite{Paul53,Myers13}. The response of the ions can be detected by counting resonantly ejected particles, by resonance absorption of the driving field \cite{Dehm62}, or through image currents in the trap electrodes \cite{Dehm68}. For certain atomic ions, it is possible to detect the ion motion through changes of scattered light that depends on velocity via the Doppler effect \cite{Wine83}. An important practical advantage of a tickle is that it only interacts with the charge of the ion, therefore it is immune to magnetic field or AC-Stark shifts that may restrict how well the HO frequency can be determined spectroscopically, for example by resolving the motional sidebands of internal transitions of ions \cite{Leib03b}. Many other implementations of HO systems have analogous mechanisms available, for example excitation of a micro-fabricated resonator by driving it with a piezo-electric element or with a capacitively coupled electric circuit.\\
\\
The tickle method can be further refined with atomic ions that are cooled close to the ground state of their motion and can be coupled to a two-level system through resolved sideband transitions \cite{Wine79,Leib03b}. Near the ground state of motion, the probability of driving a ``red sideband'' transition, where the internal state change of the ion is accompanied by reducing the number of quanta in the motional state $\ket n \rightarrow \ket{n-1}$, is strongly suppressed and can be used to determine the average harmonic oscillator occupation number $\bar{n}$ \cite{Died89,Leib03b}. Starting near the ground state, a resonant tickle can add quanta of motion such that the red sideband can be driven again, as discussed qualitatively in \cite{Home11}.
For weak excitation, $\bar{n} \leq 1$, we observe responses close to the Fourier limit of the tickle pulse, as we will describe in more detail and experimentally demonstrate.\\
\\
If the tickle excitation acts longer or with a larger amplitude, an ion in the ground state can be displaced to coherent states with an average HO occupation number $\bar{n} \gg 1$. The Rabi frequency of sideband transitions depends non-linearly on $\bar{n}$, which leads to collapse and revival of internal state changes that are one of the hallmarks of the Jaynes-Cummings model \cite{Jayn63,vonFoer74,Meek96}. Here we examine the probability of changing the internal state theoretically and experimentally, as a function of tickle detuning relative to the frequency of a HO motional mode of the ion. When probing the red sideband transition after displacing to $\bar{n} \gg 1$, we observe rich sets of features with steep and narrow side lobes around the resonance center. Such nonlinear responses can in principle be used to find the frequency of the motion with better signal-to-noise ratio than what the Fourier limit implies for smaller coherent states where $\bar{n} \leq 1$ and the response of the ion is essentially linear.\\
\\  
A sequence of coherent displacements alternating with free evolution of the motion, inspired by spin-echos \cite{Hahn50} and dynamical decoupling \cite{Erns87,Alva11}, can be used to obtain a frequency-filtered response of the ion. We implement and characterize such sequences by observing and modeling the ion response to deliberately applied, monochromatic modulations to the trapping potential curvature. Similar sequences can then be used without applied modulations. In this case, the response of the ion can be attributed to HO frequency noise that is intrinsic to our system, allowing us to characterize noise on the trap potential in a frequency range of 500 Hz to 400 kHz, a wide frequency range that has not been studied in detail in previous work \cite{Talu16}. With this method, several narrow band technical noise components (spurs) in our setup could be identified through the direct response of the ion. The noise was traced back to digital-to-analog converters (DACs) used in our setup and eliminated by replacing them with analog bench power supplies. 
\section{States and ion fluorescence signals from coherent displacements}
We consider a single ion with charge $q$ and mass $m$ confined in a harmonic trapping potential with minimum position at $\bf{r}_0$, such that the motion of the ion can be described by three normal HO modes with frequencies $\omega_x\leq \omega_y \leq \omega_z$. By using a coordinate system where the axis directions coincide with the normal mode directions, we can write the ion position as $\bf{r}=\bf{r}_0+\delta\bf{r}$. The interaction of the ion with an additional uniform electric field $\bf{E}$ can be described as 
\begin{equation}\label{Eq:TicFie}
H_{\rm{E}}=q(\bf{E}\cdot \delta\bf{r}).
\end{equation}
For the HO in the $x$-direction we introduce ladder operators $\hat{a}$ and $\hat{a}^\dag$ to write the energy as 
\begin{equation}
H_0=\hbar \omega_x \hat{a}^\dag \hat{a}.
\end{equation}
We have suppressed the ground state energy since it is a constant term that does not change the dynamics. We replace $\delta x$ by its equivalent quantum mechanical operator $\delta \hat{x} =x_0(\hat{a}^\dag+\hat{a})$ with $x_0=\sqrt{\hbar/(2 m \omega_x)}$ the ground state extent of the oscillator. For the normal mode in this direction, and in the interaction picture relative to $H_0$ the interaction with an oscillating electric field  $E_x(t)=E_0\cos(\omega t + \phi)$ becomes 
\begin{eqnarray}
H_{I}&=& \hbar \Omega_x (\hat{a}^\dag e^{i \omega_x t}+\hat{a} e^{-i \omega_x t})(e^{i (\omega t+\phi)}+e^{-i (\omega t+\phi)})\nonumber\\
&=&\hbar \Omega_x (\hat{a}^\dag e^{-i(\delta t+\phi)}+\hat{a} e^{i (\delta t+\phi)}+\hat{a}^\dag e^{i(\sigma t+\phi)}+\hat{a} e^{-i (\sigma t+\phi)}),
\end{eqnarray}
with the coupling $\Omega_x= q E_0 x_0/(2 \hbar)$ and $\delta=\omega-\omega_x$, $\sigma=\omega_x+\omega$. If the oscillating field is close to resonance with the normal mode, $|\delta| \ll \sigma$, the faster-rotating terms containing $\sigma$ can be neglected to a good approximation and the interaction takes the form of a coherent drive detuned by $\delta$
\begin{equation}
H_I \simeq \hbar \Omega_x (\hat{a}^\dag e^{-i(\delta t+\phi)}+\hat{a} e^{i (\delta t+\phi)}). 
\end{equation}
We can formally integrate the equation of motion for $H_I$ \cite{Carr65,Glaub63,Leib03} to connect an initial state $\ket{\Psi(0)}$ at $t=0$ when the electric field is switched on to the coherently displaced state after evolution for duration $t$, $\ket{\Psi(t)}$

\begin{equation}
\ket{\Psi(t)}=\hat{D}(\alpha(t))e^{i \Phi(t)}\ket{\Psi(0)},
\end{equation}
where 
\begin{eqnarray}\label{Eq:EvoSta}
\hat{D}(\alpha)&=&\exp(\alpha \hat{a}^\dag + \alpha^* \hat{a}),\nonumber\\
\alpha(t)&=&\Omega_x e^{i \phi}\int_{0}^t e^{i \delta \tau} d\tau= i \Omega_x e^{i \phi}\frac{1-e^{i \delta t}}{\delta},\nonumber\\
\Phi(t)&=&{\rm Im}\left[ \int_0^t \alpha(\tau) \{\partial_{\tau} \alpha^*(\tau)\}d\tau\right]=\left(\frac{\Omega_x}{\delta} \right)^2 [\sin(\delta t)-\delta t].
\end{eqnarray}
The phase $\Phi(t)$ can play an important role, for example, in two-qubit gates \cite{Leib03,Lee05} or interferometric experiments that combine internal degrees of freedom of the ion with motional states \cite{Monr96,Hemp13}. Here, we will be interested only in the average occupation number $\bar{n}=\langle \Psi(t)|\hat{a}^{\dag} \hat{a}|\Psi(t)\rangle$, which does not depend on $\Phi(t)$. If the initial state is the harmonic oscillator ground state, $\ket{\Psi(0)}=\ket{0}$ the average occupation is
\begin{equation}\label{Eq:Nbar}
\bar{n}(t)=|\alpha(t)|^2 =2 \left( \frac{\Omega_x}{\delta}\right)^2 [1-\cos(\delta t)].
\end{equation}
On resonance ($\delta=0$) the coherent state amplitude grows linearly in $t$ as $\alpha(t)=e^{i \phi}\Omega_x t$ and the energy of the oscillator quadratically as $\bar{n}(t)=\Omega_x^2 t^2$. For a coherent state, the probability distribution over number states $\ket{m}$ is a Poisson distribution with average $\bar{n}$
\begin{equation}
P^{(0)}_m = \frac{\bar{n}^m e^{-\bar{n}}}{m!}.
\end{equation}
An initial number state with $\ket{\Psi(0)}=\ket n$, displaced by $\hat{D}({\alpha_d})$, results in a more involved probability distribution \cite{Carr65}
\begin{equation}
P^{(n)}_m =\bar{n}^{|n-m|}e^{-\bar{n}} \frac{n_{<}!}{n_{>}!} (L_{n_<}^{|n-m|}(\bar{n}))^2,
\end{equation}
where $\bar{n}=|\alpha_d|^2$, $n_<$ ($n_>$) is the lesser (greater) of the integers $n$ and $m$ and $L^a_{n}(x)$ is a generalized Laguerre polynomial.\\
\\
In our experiments, the HO motion is coupled by laser fields to a two-level system with states labelled $\ket \downarrow$ and $\ket\uparrow$ with energy difference $E_\uparrow - E_\downarrow =\hbar \omega_0 > 0$. The internal state is initialized in $\ket\downarrow$ by optical pumping. After the state of motion is prepared, the state $\ket\downarrow \ket{\Psi(t)}$ can be driven on a red sideband, resulting in population transfer $\ket\downarrow\ket m \leftrightarrow \ket\uparrow\ket{m-1}$ for all $m>0$ while the state $\ket\downarrow\ket 0$ is unaffected \cite{Leib03b}. The Rabi frequencies depend on $m>0$ as
\begin{equation}\label{Rabif}
\Omega_{m,m-1} = \Omega_0 e^{-\eta^2/2} \eta \sqrt{\frac{1}{m}} L_{m-1}^1(\eta^2), 
\end{equation}
where $\Omega_0$ is the Rabi frequency of a carrier transition $\ket\downarrow \leftrightarrow \ket\uparrow$ of an atom at rest, $\eta=k_x x_0$ is the Lamb-Dicke parameter, with $k_x$ the component of the effective wavevector along the direction of oscillation and $L^{\beta}_m(x)$ is a generalized Laguerre polynomial \cite{Leib03b}. After driving the red sideband of state $\ket\downarrow\ket{\Psi(t)}$ for duration $\tau$, the probability of having flipped the internal state to $\ket\uparrow$ is
\begin{equation}
P_{\uparrow}(\tau) = \frac{1}{2}\left[1-P^{(n)}_0-\sum_{m=1}^\infty P^{(n)}_m \cos(2 \Omega_{m.m-1} \tau)\right]. 
\end{equation}
We calibrate the red sideband drive time to be equivalent to a $\pi$-pulse on the $\ket\downarrow \ket{1} \leftrightarrow \ket\uparrow \ket{0}$ transition, which implies $2 \Omega_{1,0} \tau = \pi$. For an arbitrary displaced number state the probability of the ion to be in $\ket\uparrow$ becomes
\begin{equation}\label{Eq:ProbUp}
P^{\pi}_{\uparrow} = \frac{1}{2}\left[1-P^{(n)}_0-\sum_{m=1}^\infty P^{(n)}_m \cos(\pi\frac{\Omega_{m.m-1}}{\Omega_{1,0}})\right]. 
\end{equation}
This probability is not a monotonic function of $\bar{n}$ and exhibits maxima and minima as the displacement changes. Experimental observations of this behavior for displaced number states and comparisons to the predictions of Eq. (\ref{Eq:ProbUp}) will be discussed in section \ref{Sec:ExpRes}. When the detuning in Eq. (\ref{Eq:Nbar}) is $\delta \neq 0$, the coherent drive displaces $\ket{\Psi(0)}$ along circular trajectories in phase space that can turn back onto themselves. Every time $\delta~ t= m \pi$ with $m$ a non-zero integer, $|\alpha(t)|$ will reach a maximum of $2 \Omega_x/\delta$ for $m$ odd and return to zero for $m$ even. The non-monotonic behavior of $P^{\pi}_{\uparrow}$ with respect to $\bar{n}$ creates feature-rich lineshapes when this probability is probed as a function of the displacement detuning $\delta$ relative to the HO frequency. 
\section{Noise sensing with motion-echo sequences}
The motion displacements discussed above enable sensitive tests of the ion's motional coherence. Intrinsic noise at the motional frequencies heats the ion out of the ground state, and is observed in all traps at a level that often exceeds resistive heating by orders of magnitude. This ``anomalous heating,'' is well documented \cite{Turch00,Brownnutt15}, but the sources are not well understood. On much longer time scales than the ion-oscillation period, motional frequencies are known to drift over minutes and hours due to various causes, for example slow changes in stray electric fields and drifts of the sources that provide the potentials applied to the trap electrodes. Much less work has been done to characterize noise in the frequency range in between the HO frequency and slow drift \cite{Talu16}. Here, we construct sequences of coherent displacements that alternate with periods of free evolution and suppress the sensitivity to slow drifts of the harmonic oscillator frequency. This allows us to isolate noise in a specific frequency band, in analogy to an AC-coupled electronic spectrum analyzer. Coherent displacements can be implemented on time scales of order of a few $\mu$s, making this method suitable for detecting noise at frequencies in a range of 500 Hz to 400 kHz in the experiments described here. 
\subsection{Basic principle}
The sequences of coherent displacements discussed here are closely related to spin-echo experiments and dynamical decoupling in two-level systems \cite{Hahn50,Erns87,Alva11}. In analogy to the classic\\ \centerline
{($\pi/2$-pulse)-$\tau_a$-($\pi$-pulse)-$\tau_a$-($\pi/2$-pulse)}\\
spin-echo sequence \cite{Hahn50} with $\tau_a$ the duration of a free-precession period, the ideal ``motion-echo''pulse sequence consists of\\
\centerline {$\hat{D}(\Omega_x \tau_d/2)$-$\tau_a$-$\hat{D}(-\Omega_x \tau_d)$-$\tau_a$-$\hat{D}(\Omega_x \tau_d/2)$,}\\  
where the minus sign in the argument of the second displacement indicates that the phase $\phi$ of the displacement drive has changed by $\pi$ relative to the other displacement operations. To simplify this initial discussion, we assume that all displacements are instantaneous and not affected by fluctuations in the oscillator frequency. This condition is similar to the ``hard-pulse'' limit for spin-echo sequences.
\begin{figure}
    \centering
    \includegraphics[width=\textwidth]{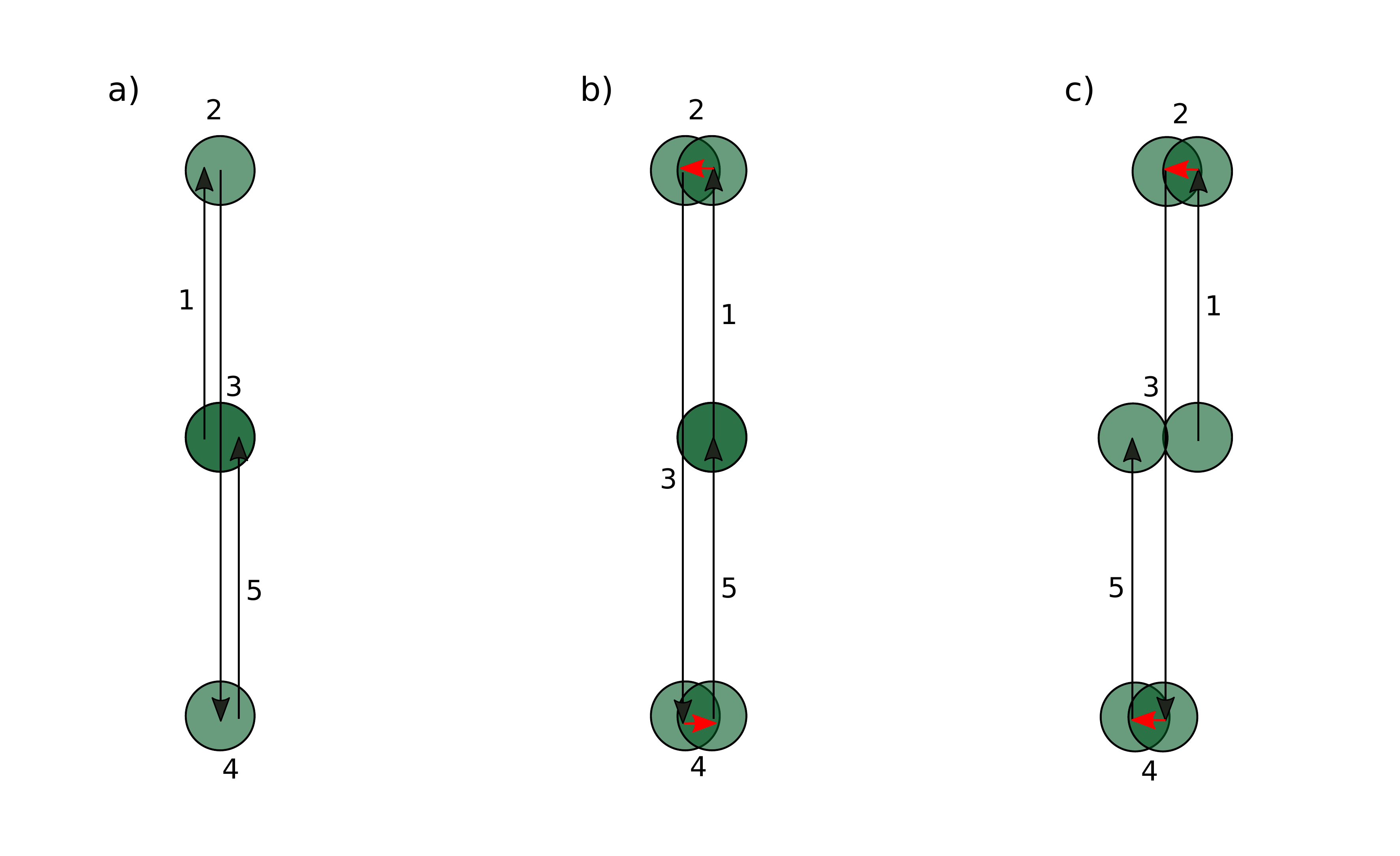}
    \caption{\label{Fig:IntDis} Schematic phase-space sketch of the displacements in the simplest motion-echo sequence. Here, all displacements are assumed to act instantaneously (hard-pulse limit), such that the effect of oscillator detuning during displacements can be neglected. (a) Without fluctuations of the oscillator frequency, the ground-state minimum uncertainty disk (green) is coherently displaced by $\Omega_x \tau_d/2$ (step 1), then remains stationary during a free-precession period (step 2), it is then displaced symmetrically through the origin by $-\Omega_x \tau_d$ (step 3), followed by another free-precession period (step 4). The final displacement by $\Omega_x \tau_d/2$ (step 5) returns the state to the origin. (b) With a small, constant detuning, the state drifts perpendicular to the direction of the first displacement in step 2. However, it drifts an equal amount in the opposite direction during step 4, to still end up in the ground state after step 5. This immunity to constant detuning can be thought of as a HO analogy to a spin-echo sequence in a spin-1/2 system. (c) If the detuning changes sign between steps 2 and 4, the state does not return to the origin. In all three cases (a)-(c), the final state reflects the sum of additional displacements during the operation of the sequence that are caused by time-dependent changes in HO detuning.}
\end{figure}
If the frequency of the oscillating electric field in Eq. (\ref{Eq:TicFie}), which we call the ``local oscillator frequency'' in this context, is on resonance with the HO frequency, the displacements in the sequence add up to zero, so any initial state is displaced back onto itself at the end of the sequence (see Fig. \ref{Fig:IntDis} (a)). In analogy to a spin-echo sequence, if the local oscillator differs from the HO frequency by a small, constant detuning $\delta \ll 2 \pi/\tau_d$, the sequence will still result in a final state that is very close to the initial state (see Fig. \ref{Fig:IntDis} (b)). However, if the detuning changes sign between free-precession periods (see Fig. \ref{Fig:IntDis} (c)), the final state will not return to the initial position and in general information about the oscillator frequency fluctuations can be gained from the final displacement. This basic echo sequence can be expanded by including additional blocks of the form\\
\centerline {$\tau_a$-$\hat{D}(\Omega_x \tau_d)$-$\tau_a$-$\hat{D}(-\Omega_x \tau_d)$}\\
after the first displacement $\hat{D}(-\Omega_x \tau_d)$ in analogy to dynamical decoupling sequences in two-level systems. Ideally, this increases the number of free-precession sampling windows which leads to a longer sampling time and narrower filter bandwidth of the extended sequence, while still producing no total displacement from the initial state if the HO is stable, even if the local oscillator is slightly detuned from the HO resonance.
\subsection{Effects of oscillator frequency fluctuations} 
If the local oscillator frequency is not on resonance with the HO, or if the detuning is not constant in time, a realistic coherent drive (not assuming the hard-pulse limit) will not always displace the state of motion along a straight line. The differential equation describing the coherent displacement $\alpha(t)$ as a function of time in that generalized case is (from Eq. (\ref{Eq:EvoSta}))
\begin{equation}\label{Eq:DifEqu}
    \dot{\alpha}(t)=\alpha i \delta(t)+\Omega_x e^{i \phi},
\end{equation}
where $\delta(t)$ is the instantaneous detuning between the HO and the local oscillator at time $t$. In the special case where $\delta$ does not depend on time and $\alpha(0)=0$, the solution is $\alpha(t)$ from Eq. (\ref{Eq:EvoSta}). If there is noise on the trap frequency, $\delta(t)$ will fluctuate randomly as a function of $t$. A general solution of Eq. (\ref{Eq:DifEqu}), at time $t_0+\tau$ as it evolves from the initial state $\alpha(t_0)$ at time $t_0$, can be formally written as 
\begin{equation}\label{Eq:ForSol1}
\alpha(t_0,\tau)=\exp[i I_1(t_0,\tau)][\alpha(t_0)+\Omega_x e^{i \phi} I_2(t_0,\tau)],    
\end{equation}
with
\begin{eqnarray}\label{Eq:ForSol2}
I_1(t_0,\tau)&=&\int_{t_0}^{t_0+\tau} \delta(\tau_1) d \tau_1, \nonumber \\
I_2(t_0,\tau)&=&\int_{t_0}^{t_0+\tau} \exp\left[-i \int_{t_0}^{\tau_2} \delta(\tau_1) d\tau_1 \right] d \tau_2.
\end{eqnarray}
This form is useful for numerical calculations and can be explicitly solved for special cases of $\delta(t)$. Motion-echo sequences are most useful if the  accumulation of phase during $\tau$ is small, $I_1(t_0,\tau)\ll 2 \pi$. In such cases, we can expand the exponential functions in Eqs. (\ref{Eq:ForSol1}) and (\ref{Eq:ForSol2}) to linear order and approximate
\begin{eqnarray}\label{Eq:PhaInt}
\alpha(t_0,\tau)&\simeq&\{\alpha(t_0)+\Omega_x \tau e^{i \phi}\}+i(\alpha(t_0)+\Omega_x \tau e^{i \phi}) I_1(t_0,\tau)\nonumber \\ & &-i \Omega_x e^{i \phi} I_3(t+0,\tau), \nonumber \\
I_3(t_0,\tau)&=&\int_{t_0}^{t_0+\tau} \left( \int_{t_0}^{\tau_2} \delta(\tau_1) d\tau_1 \right) d  \tau_2.
\end{eqnarray}
The different terms in Eq. (\ref{Eq:PhaInt}) have straightforward interpretations: the term in curly braces characterizes the displaced coherent state for no detuning, $\delta(t)=0$. Finite detuning rotates this state around the origin in phase space and to lowest order this effect is captured by the term proportional to $I_1(t_0,\tau)$. The final term reflects the effect of the detuning while the state is displaced by a coherent drive, which results in a correction proportional to $I_3(t_0,\tau)$. For a free-precession period, $\Omega_x=0$, during displacement by a coherent drive, $\Omega_x \neq 0$.\\
\\
In this linear approximation, it is straightforward to keep track of the displacements and the corrections from $\delta(t)\neq0$ when periods of driving and free precession are concatenated. Because corrections on earlier corrections are higher order than linear, the correction from each period only acts on the zero-order displacement of any previous period. This implies that the zero-order terms in curly brackets and the corrections can be summed up separately for a sequence. In this way, we can calculate the total zero-order displacement $\alpha_n$ and first order correction $\Delta \alpha_n$ of a sequence with $n$ steps starting at time $t=0$ in state $\ket{\alpha(0)}$. For the $k$-th step starting at $t_k$, the displacement drive Rabi frequency is $\Omega_{x,k}$, the drive duration $\tau_k$, and the phase $\phi_k$. In the linear approximation with $\alpha(0)=\alpha_0$ the sums are
\begin{eqnarray}\label{Eq:FinDis}
\alpha_n&=&\alpha_0+\sum_{k=1}^n \Omega_{x,k} \tau_k e^{i \phi_k},\nonumber\\
\Delta \alpha_n &=& i \sum_{k=1}^n \left\{ \left(\alpha_{k-1}+\Omega_{x,k} \tau_k e^{i \phi_k}\right)I_1(t_k,\tau_k)-\Omega_{x,k} e^{i \phi_k}I_3(t_k,\tau_k) \right\}.
\end{eqnarray}
\subsection{Motion-echo sequences}
We restrict ourselves to sequences with $N$ steps acting on an initial ground state, $\alpha_0=0$, where the sum over all unperturbed displacements of a sequence is $\alpha_N=0$. In this way, the final state is equal to $\ket{\Delta \alpha_N}$ and directly reflects the effects of non-zero detuning. Moreover, we can construct the displacements in such a way, that a constant detuning $\delta\neq0$, results in $\Delta \alpha_N=0$. This mimics the feature of spin-echo sequences that small constant detunings have no effect on their final state. The motion-echo sequences preserve this feature, if the linear approximation is valid, even when taking the effect of the detuning onto the displacement operations into account. For constant $\delta$, the integrals $I_1(t_k,\tau_k)=\delta \tau_k$ and $I_3(t_k,\tau_k)=1/2~\delta \tau_k^2$ are independent of $t_k$ and the total displacement simplifies to
\begin{equation}\label{Eq:StaSum}
\Delta \alpha_N = i \delta \sum_{k=1}^N \tau_k (\alpha_{k-1} + 1/2~\Omega_{x,k} e^{i \phi_k} \tau_k).
\end{equation}
For the motion-echo sequences, $\Omega_{x,k}=\Omega_x$ is the same for all displacements and the coherent state parameter before each of the $\hat{D}(\pm \Omega_x \tau_d)$ operations is $\mp \Omega_x \tau_d/2$. In this case, the second contribution in the $(...)$ braces is $\pm 1/2 \Omega_x \tau_d$, equal and opposite to the initial state parameter, so all displacement terms in the sum Eq. (\ref{Eq:StaSum}) are equal to zero individually, except for the first and last displacement which is $\hat{D}(\Omega_x \tau_d/2)$. However, since $\alpha_0=0$, these terms sum to $i \delta \tau_d/2 (1/2 \Omega_x \tau_d/2 -\Omega_x \tau_d/2 +1/2 \Omega_x \tau_d/2)=0$, which leaves only the free precession terms to be considered. All sequences contain an even number $2 n_a$ of free-precession periods ($n_a>0$, integer), with half of them contributing $i \delta \tau_a \Omega_x \tau_d/2$ each and the other half $-i \delta \tau_a \Omega_x \tau_d/2$, so, as previously noted (see Fig. \ref{Fig:IntDis}) these terms also sum to zero and $\Delta \alpha_N=0$ for a constant detuning.
\subsection{Response to a monochromatic modulation}
Next, we can determine the response to a monochromatic modulation at frequency $\omega_n$ of the form $\delta_n(t)=A_n \cos(\omega_n t+\phi_{n})$. On the one hand, the HO frequency can be deliberately modulated in this way, which enables us to compare the response of the motion-echo sequence to the theoretical expectation. On the other hand, some of the frequency noise acting on the oscillator can be characterized as a noise spectrum consisting of a sum of such modulation terms with distinct frequencies $\omega_{n,j}$, possibly varying amplitudes $A_{n,j}$ and random phases $\phi_{n,j}$. In addition, the harmonic oscillator may be affected by noise with a continuous spectrum, but we will restrict ourselves to discrete, narrow-band noise spurs here. The noise spectrum can be characterized with motion-echo sequences, if the response to  a monochromatic modulation at $\omega_n$ allows for determination of that frequency within a band that depends on the resolution of the sequence. The amplitude of the response $\Delta \alpha_N$ is proportional to the noise amplitude and is zero when averaged over the random noise phase $\phi_n$ (denoted by $\langle...\rangle$), but because the final occupation $\bar{n}_{\rm fin}$ is proportional to $|\alpha_N|^2$, after integrating over $\phi_n$, we get an average final occupation
\begin{equation}
\langle\bar{n}_{\rm fin}\rangle =\frac{1}{2 \pi} \int_0^{2 \pi} |\Delta \alpha_N|^2 d\phi_n.
\end{equation}
This is proportional to the noise power inside the filter bandwidth of the motion-echo.\\
\\
For the monochromatic modulation, the integrals $I_1$ and $I_3$ have analytic solutions:
\begin{eqnarray}
I_1(t_0, \tau)&=&\frac{A_n}{\omega_n}[\sin(\omega_n(t_0+\tau)+\phi_{n})-\sin(\omega_n t_0+\phi_{n})] \nonumber\\
    I_3(t_0,\tau)&=&\frac{A_n}{\omega_n^2}[\cos(\omega_n t_0+\phi_{n})-\cos(\omega_n(t_0+\tau)+\phi_{n})-\nonumber\\
    & &-\omega_n \tau \sin(\omega_n t_0+\phi_{n})].
\end{eqnarray}
Now, the integrals depend on $t_0$ and $\tau$, therefore the sum over a motion-echo sequence is non-zero in general. Inserting the integrals into Eq. (\ref{Eq:FinDis}) and summing over the motion-echo sequences is straightforward but tedious, and yields closed expressions for the final displacement $\Delta \alpha_N$ and the corresponding average occupation number of the motion $\bar{n}_{\rm fin}=|\Delta \alpha_N|^2$.
Taking the average over the random phase $\phi_n$ yields
\begin{eqnarray}\label{Eq:nfin}
    \langle \bar{n}_{\rm fin} \rangle &=& \frac{8 A_n^2 \Omega_x^2}{\omega_n^4}
   \sin^2[\omega_n \tau_d/4]\{\cos[\omega_n \tau_d/4]-\cos[\omega_n(\tau_a+ 3 \tau_d/4)]\}^2\times\nonumber\\
    & &\times \frac{\sin^2[n_{\rm a}\omega_n (\tau_d+\tau_a)]}{\sin^2[\omega_n(\tau_d+\tau_a)]}.
\end{eqnarray}
If the free-evolution time $\tau_a$ is varied in the motion-echo sequence, the expression in the upper line produces an envelope that is oscillating at frequency $\omega_n$ with phase shifts proportional to $\tau_d$.\footnote{The expression in the lower line is equivalent to the intensity far-field pattern of a transmission grating with $n_a$ slits \cite{Born80} and describes a $n_a$-times sharper response that produces a more narrowly peaked interference pattern with nearly symmetric side lobes. } The first main peak appears when $\omega_n (\tau_d+\tau_a)\simeq \pi$ and the spacing between adjacent main peaks is exactly $\Delta \tau_a = 2\pi/\omega_n$, which allows for determination of $\omega_n$ from this interference pattern.  The width of the narrow main peaks can be characterized by the distance $\delta \tau_a$ of the two zeros of the response closest to a peak, which are spaced by $\delta \tau_a =2 \pi/(n_a \omega_n)$. It is possible to resolve a pair of main peaks produced by modulations at $\omega_n$ and $\omega_n +\delta \omega_n$ respectively, as separate maxima if $|\delta \omega_n| \geq \pi/[n_a(\tau_a+\tau_d)]$. If a continuous noise power spectral density $a_n^2(\omega_n)$ is sampled in this way, $\delta \omega_n$ determines the bandwidth of the sample filter that relates the noise power density to the actual noise power detected in this band.\\
\\
To have $\langle \bar{n}_{\rm fin} \rangle$ approximately reflected in the ion state population $P^\pi_{\uparrow}$, the average mode occupation should be kept below $\langle \bar{n}_{\rm fin} \rangle \leq 1$, which is possible by reasonable choices for the displacement $\Omega_x \tau_d$ and the number of free-precession periods $2n_a$. Choosing quantities that are too large has the same effect as over-driving the mixer in an electronic spectrum analyzer, which leads to a response that is not linear in the input signal, resulting in a distorted output.
\section{Experimental implementation and results}
\subsection{Experimental setup}
Experiments were performed with a single $^9 $Be$ ^+$ ion trapped 40 $\mu$m above a cryogenic ($\simeq$ 4 K) linear surface-electrode trap described elsewhere \cite{Brown11,Wils14}. The coherent displacements are performed on the lowest frequency mode (axial) of the three orthogonal harmonic oscillator modes of the ion, with frequency  $\omega_x \simeq 2\pi \times 8$ MHz. We use two levels within the electronic $^2 S_{1/2}$ ground-state hyperfine manifold, $\ket{F=1,m_F = -1} = \ket \uparrow$, $\ket{F=2,m_F=-2} = \ket{\downarrow}$, where $F$ is the total angular momentum and $m_F$ is the component along the quantization axis, defined by a 1.43\,mT static magnetic field. Direct ``carrier''-transitions between the states $\ket \downarrow \ket n$ and $\ket \uparrow \ket n$ are driven by microwave fields induced by a $\omega_0 \simeq 2\pi \times 1.281$ GHz current through one of the surface trap electrodes. \\
\\
The ion is prepared in $\ket{\downarrow}\ket{0}$ with a fidelity exceeding 0.99 by Doppler laser cooling, followed by ground-state cooling \cite{Leib03} and optical pumping. Sideband transitions $\ket \downarrow \ket n \leftrightarrow \ket \uparrow\ket {n\pm 1}$ are implemented with stimulated Raman transitions driven by two laser fields that are detuned from the $^{2}S_{1/2} \rightarrow $ $^{2}P_{1/2}$ transition ($\lambda \simeq 313$ nm) by approximately 40 GHz \cite{Monr95}. This allows us to prepare nearly pure number states of the motion as described in more detail in \cite{McCormick18}. We implement the tickle by applying a square-envelope pulse with oscillation frequency near the axial mode frequency to the same electrode that is used for the microwave-driven hyperfine transitions, which produces an electric field at the position of the ion with a component along the direction of the motional mode.\\
\\
We distinguish measurements of the $\ket{\uparrow}$ and $\ket{\downarrow}$ states with state-dependent fluorescence \cite{Leib03}. When scattering light from a laser beam resonant with the $\ket{\downarrow} \leftrightarrow \ket{P_{3/2}, F=3,m_F=-3}$ cycling transition, 11 to 13 photons are detected on average over 400 $\mu$s with a photo-multiplier if the ion is in $\ket{\downarrow}$, while only 0.2 to 0.5 photons (dark counts and stray light) are detected on average if the ion is projected into $\ket{ \uparrow}$. \\
\\
In the experiments detailed below, the signal indicates the deviation of the final motional state from $\ket{n=0}$. Population in the ground state of motion is discriminated from that in excited states of motion by performing the RSB pulse theoretically described in Eq. (\ref{Eq:ProbUp}), connecting population in $\ket\downarrow \ket{n>0}$ to $\ket{\uparrow}\ket{n-1}$ while leaving population in $\ket\downarrow\ket{n=0}$ unchanged. For average excitation $\bar{n} \leq 1/2$ the probability $P_{\uparrow}^\pi$ of changing the internal state is approximately equal to $\bar{n}$. A subsequent microwave carrier $\pi$-pulse exchanges population in $\ket{\downarrow} $ and $\ket{\uparrow}$. The latter state has a low average count rate, which minimizes shot noise in the photomultiplier signal. This is helpful when determining small deviations from $\ket{n=0}$ with high signal-to-noise ratio.  \\
\subsection{Displaced number states} \label{Sec:ExpRes}
\begin{figure}
\centering
\includegraphics[width = 0.9 \textwidth]{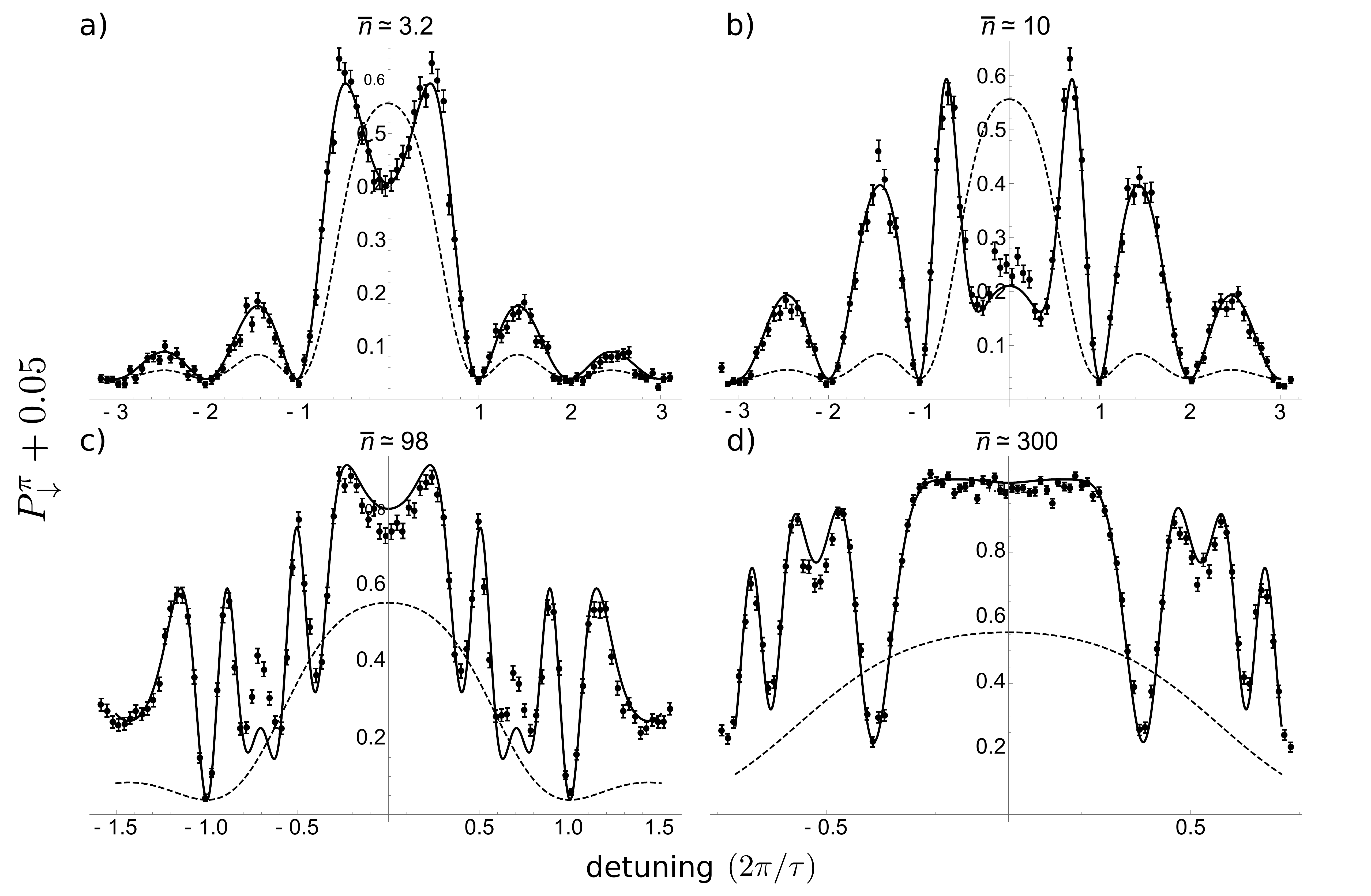}
\caption{\label{Fig:lineshapes} Response of ion to tickle excitation versus detuning from ion oscillation frequency for maximal excitations on resonance of a) $\bar{n} = 3.22(3)$, b) $\bar{n} = 10.4(1)$, c) $\bar{n} = 98.4(7)$ and d) $\bar{n} = 299(1)$. As the average occupation of the coherent state increases, the line shape becomes more sharply featured. Solid lines are fits with free parameters $\Omega_x$ and $\omega_x$. We determine the maximum excitation $\bar{n}$ of each experiment from the fitted values of $\Omega_x$. Dashed lines represent a Fourier-limited response resulting from a weaker excitation ($\bar{n} = 0.64$) for comparison.}
\end{figure}

\begin{figure}
\centering
\includegraphics[width = \textwidth]{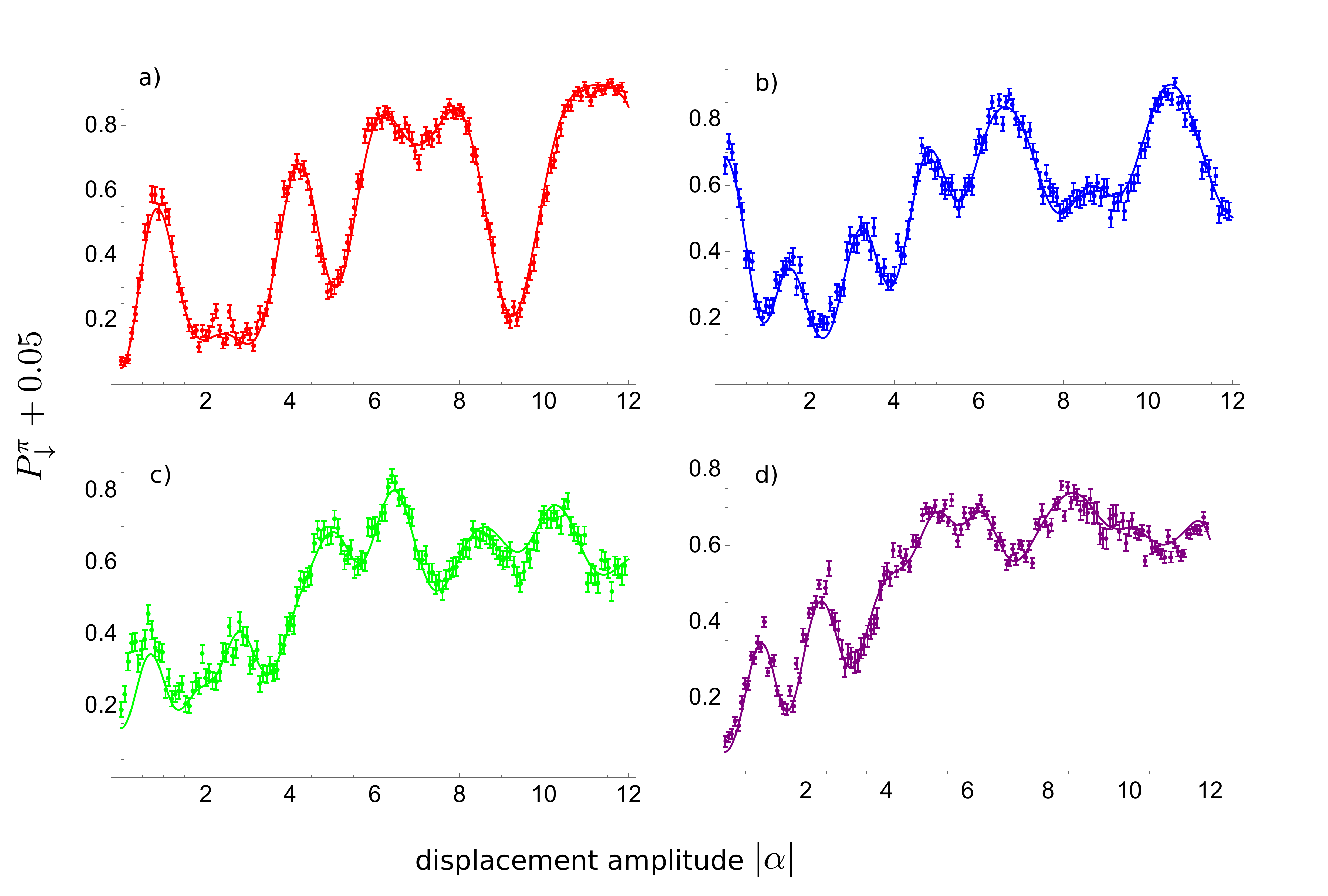}
\caption{\label{Fig:displacednumberstates} Response of ion to tickle excitation versus displacement $\alpha$ for initial number states a) $n=0$, b) $n=2$, c) $n=4$ and d) $n=6$. As the displacement of the ion's motion increases, the Rabi frequency of the RSB interaction varies non-monotonically. Lines are produced with theory in Eq. (\ref{Eq:ProbUp}) using fit parameters of the tickle strength $\Omega_x$ and contrast of the final state readout, and an experimentally determined vertical offset of 0.05(1) to account for background counts and imperfect initial state preparation. }
\end{figure}

As briefly described in \cite{Home11}, tickling an ion that has been cooled to near the motional ground state to determine the ion oscillation frequency is a practical calibration tool. The experiment is performed here as follows: the ion is cooled to near the ground state and prepared in $\ket{\downarrow}$, then a tickle tone with a fixed amplitude and detuning $\delta$ is applied for a fixed duration $\tau_d$ = 13 $\mu$s. The resulting coherent state is characterized by applying a RSB $\pi$-pulse for the $\ket \downarrow \ket 1 \rightarrow \ket \uparrow \ket0 $ transition followed by a microwave carrier $\pi$-pulse, then detected via state-selective fluorescence as described above. \\
\\
\begin{figure}
\centering
\includegraphics[width = 0.8 \textwidth]{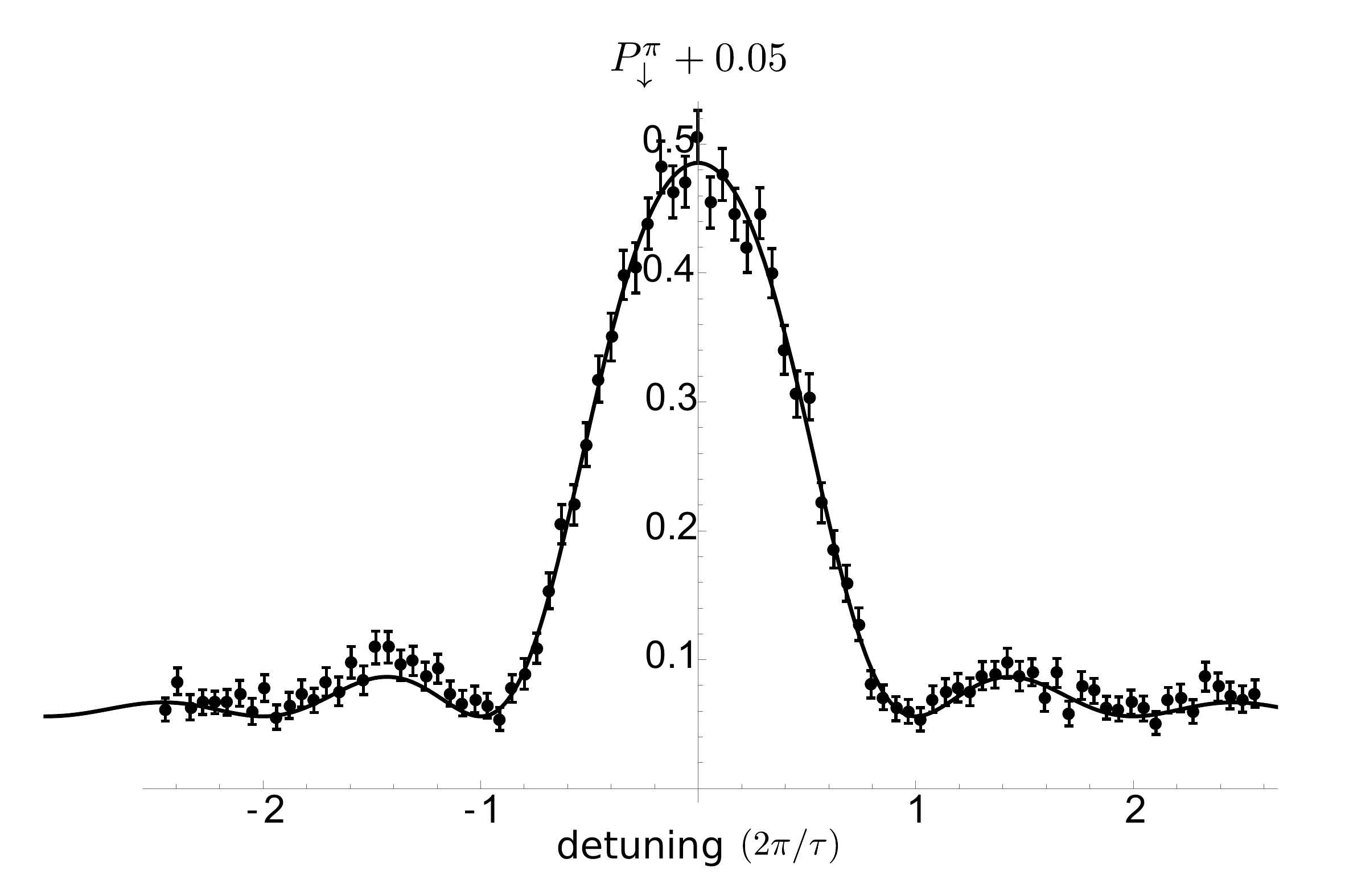}
\caption{\label{Fig:tickle} Spin-flip probability $P_\downarrow^\pi$ (See Sec. 4.2) of ion after 13 $\mu$s tickle excitation on $\ket\downarrow$ versus detuning from ion oscillation frequency. The average occupation $\bar{n}$ of the ion motion in response to tickle excitation is mapped onto the spin state by performing a RSB pulse, which connects levels $\ket{\downarrow}\ket{n}$ to $\ket\uparrow\ket{n-1} $ for $n>0$, while leaving population in $\ket\downarrow\ket{n=0}$ unchanged. A subsequent microwave carrier $\pi$-pulse exchanges populations in $\ket\uparrow$ and $\ket\downarrow$ to reduce measurement projection noise. The solid line is a fit using Eq. (\ref{Eq:Nbar}) and free parameter $\bar{n}$ and an experimentally determined vertical offset of 0.05(1) to account for background counts and imperfect ground state cooling. The fit yields an on resonance average occupation of $\bar{n}=0.61(1)$.}
\end{figure}
The symbols with error bars (1-$\sigma$ statistical error, from shot noise in the photo-multiplier count rate averaged over 600 experiments per detuning value) in Fig. \ref{Fig:tickle} show the measured $P_\downarrow^\pi$ as a function of tickle detuning for low on-resonance occupation ($\bar{n}=0.61(1)$). The line shape is well described by Eqs. (\ref{Eq:Nbar}) and (\ref{Eq:ProbUp}). The solid line is a fit to these equations with $\bar{n}=0.61(1)$ as the only free parameter after subtracting an offset of 0.05(1) due to stray light background and imperfect ground state cooling that was determined independently. In this case, $P_\downarrow^\pi$ is roughly linear in $\bar{n}$ and reflects the sinc$^2$-shape of the Fourier transform of the square-envelope tickle pulse. Keeping the excitation small gives us the practical advantage that only one prominent peak in $P_\downarrow^\pi$ versus detuning exists, making fitting to find the resonance frequency straightforward. However, the precision with which we can determine the frequency of oscillation is Fourier-limited by the time that we apply the pulse. \\
\\
With the development of a theoretical understanding of line shapes for larger excitations, where $P_\downarrow^\pi$ is non-linear in $\bar{n}$, we have found that we can determine the resonant frequency with a precision that increases approximately linearly with the size of the excitation $|\alpha|$ in the range of $0<|\alpha|<17$, which implies that we can improve on the Fourier limit of the tickle pulse. We have measured $P^\pi_\downarrow$ (see Eq. \ref{Eq:ProbUp}) versus detuning of the tickle frequency for various displacement amplitudes up to $|\alpha| \approx 17$, corresponding to a coherent state with an average occupation of $\bar{n} \approx 300$. Fig. \ref{Fig:lineshapes} shows four such cases with  $\bar{n}$ of $3.22(3)$ (Fig. \ref{Fig:lineshapes}a), $10.4(1)$ (Fig. \ref{Fig:lineshapes}b), $98.4(7)$ (Fig. \ref{Fig:lineshapes}c) and $299(1)$ (Fig. \ref{Fig:lineshapes}d). The lines are fits with free parameters $\Omega_x$ and HO resonance frequency $\omega_x$. An experimentally determined vertical offset of 0.05(1) is added to the fit function to account for background counts, as in the evaluation of the data presented in Fig. \ref{Fig:tickle}.\\
\\
The steep slopes of some of the line-shape features imply a stronger response to small changes in the detuning, as compared to cases where $\bar{n}\leq 1$ (dashed lines in Fig. \ref{Fig:lineshapes}). Moreover, the response is symmetric around $\delta=0$, so these steeper slopes can be exploited to find $\delta=0$ without a detailed understanding of the line shapes beyond this symmetry. This enables line-center determination with a signal-to-noise ratio beyond the Fourier limit of the linear response ($\bar{n}\leq 1$).\\
\\
%
We can also scan the amplitude of the coherent displacement while the tickle frequency is resonant with the ion's oscillation. As the amplitude of the coherent state increases, the Rabi frequency of the RSB interaction varies with $n$ as predicted by Eq. (\ref{Rabif}), producing the non-monotonic response of the ion's fluorescence shown in Fig. \ref{Fig:displacednumberstates}  together with fits based on Eq. (\ref{Rabif}) with $\bar{n}$ as a free parameter. We perform this experiment by preparing the ion in pure number states $n=0$ (Fig. \ref{Fig:displacednumberstates} a)), $n=2$ (Fig. \ref{Fig:displacednumberstates} b)), $n=4$ (Fig. \ref{Fig:displacednumberstates} c)) and $n=6$ ((Fig. \ref{Fig:displacednumberstates} d)) and applying a resonant tickle tone with fixed $\Omega_x$ for times ranging from $0.05 to 12~\mu$s, resulting in coherent displacements up to $|\alpha_d| \approx 17$. \\
\subsection{Motion-echo experiments}
\begin{figure}
\centering
\includegraphics[width = \textwidth]{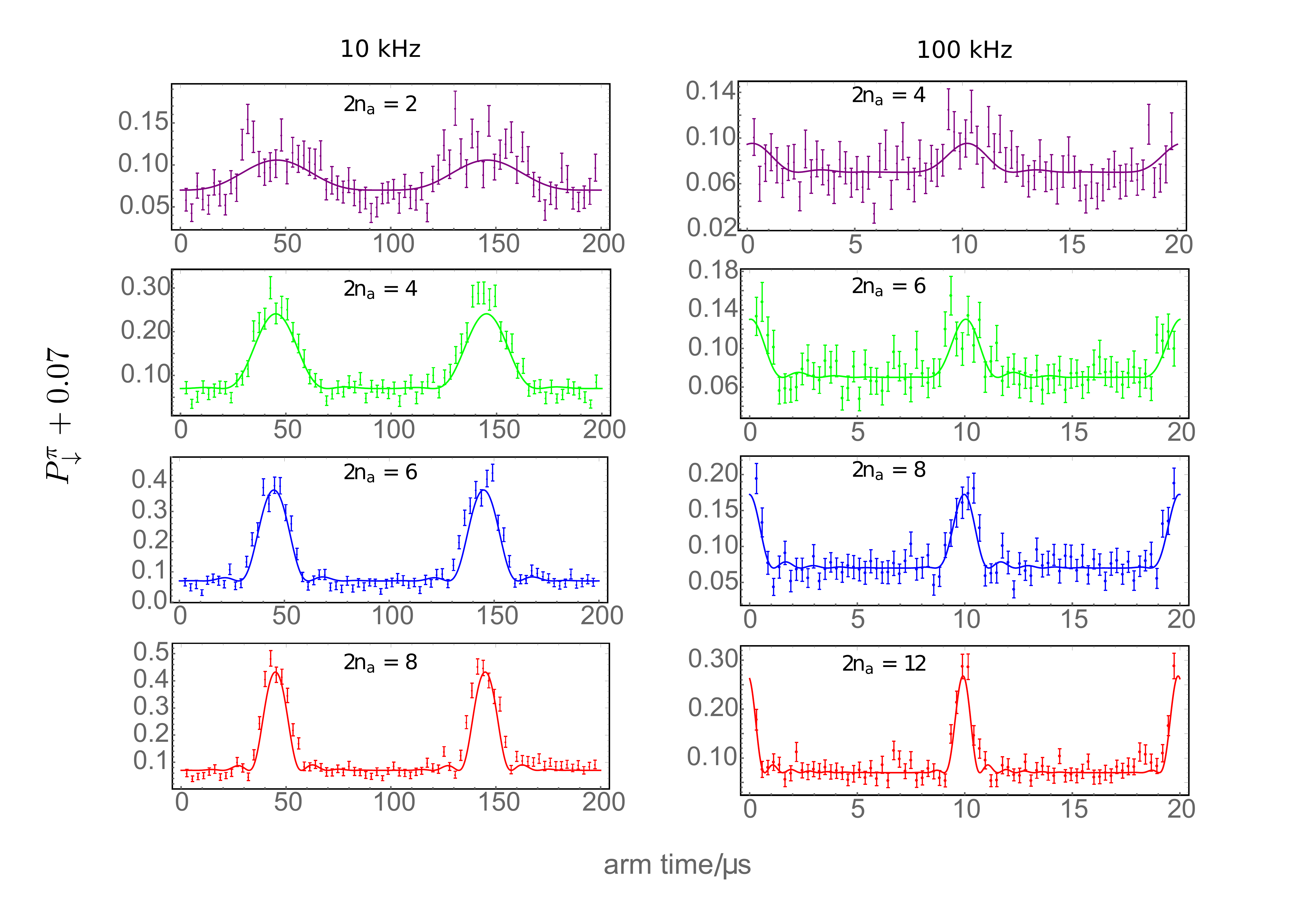}
\caption{\label{Fig:10khzarms} Characterization of motion-echo experiments with applied tones of 10 kHz and 100 kHz for various numbers of free-precession periods $2 n_a$ (see top of plots). Lines are fits based on Eqs. (\ref{Eq:nfin}) and (\ref{Eq:ProbUp}) with free parameters $A_n$. For the experiments with the 10 kHz tone applied,  fitted values of $A_n$ were $2 \pi \times$ 1.3(4) kHz, 1.5(1) kHz, 1.4(1) kHz and 1.2(1) kHz for $2n_a$ = 2, 4, 6 and 8, respectively. Likewise for the 100 kHz tone, fitted values of $A_n$ were $2 \pi \times$ 1.0(4) kHz, 1.1(3) kHz, 1.1(2) kHz and 1.1(1) kHz for $2n_a$ = 4, 6, 8 and 12, respectively.} 
\end{figure}

\begin{figure}
\centering
\includegraphics[width = \textwidth]{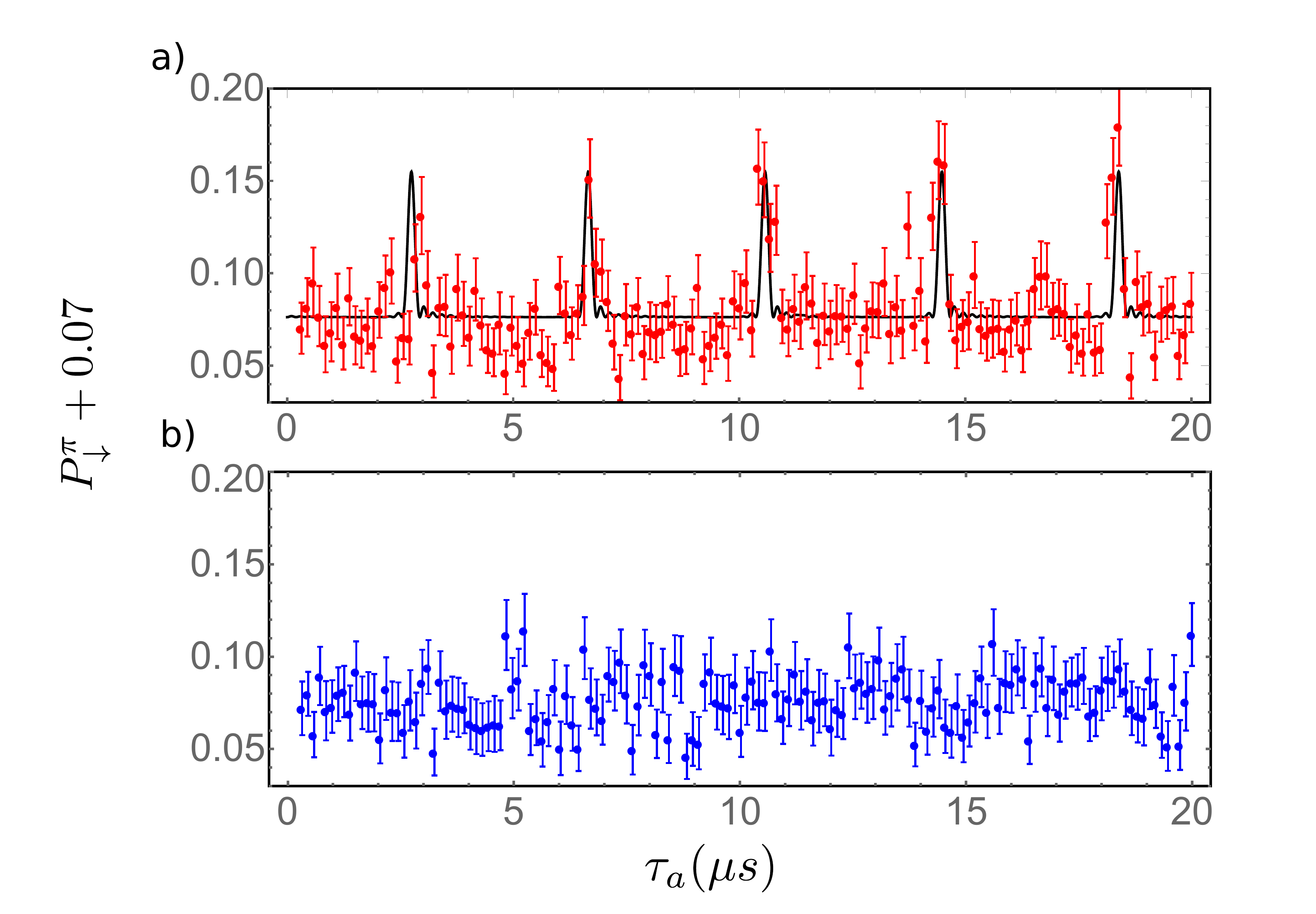}
\caption{\label{Fig:realnoise} Noise sensing with motion-echo experiments. The two sub-plots compare noise on electrode potentials delivered from two different sources: a) Digital-to-Analog converters (DACs) and b) linear bench power supply. Experiments were performed with $n_a=10$ (20 free-precession periods) and coherent displacements of $\Omega_x \tau_d /2 = 3.44(2) $. The solid black line in a) is a fit to the data taken on the DACs based on Eq. (\ref{Eq:nfin}) with free parameters $A_n$ and $\omega_n$, and a vertical offset of 0.07 to account for background counts and imperfect state preparation. From this fit, we determine that the DACs introduce noise at $\omega_n/(2 \pi) \simeq$ 260 kHz with $A_n= 2 \pi \times 2.4(2)$ kHz.} 
\end{figure}

We first perform the motion-echo experiments with a 10 kHz or 100 kHz tone applied to one of the trap electrodes. The tone modulates the potential curvature of the trap at the position of the ion and therefore the ion's oscillation frequency. The purpose of this is two-fold: To experimentally characterize the response of the ion and compare the results against theory for a known perturbation and to explore the range of noise that is detectable with this method in our setup. With the tone applied continuously and the phase $\phi_n$ changing randomly from experiment to experiment (to mimic the uncontrolled phase of noise), we perform a series of motion-echo sequences with various numbers of free-precession periods 2$n_a$, scanning the wait time $\tau_a$. We find that $P_\downarrow^\pi$ depends on the relationship between $\tau_a$ and the frequency of the applied tone in the expected manner according to Eq. (\ref{Eq:nfin}). This can be seen by comparing the measured points in Fig. \ref{Fig:10khzarms} to the solid lines that show fits with $A_n$ as a free parameter. All fitted values of $A_n$ are consistent with each other to within 2 times the standard deviation (see caption of Fig. \ref{Fig:10khzarms}) and indicate a relative modulation depth of $A_n/\omega_x \simeq 1.4 \times 10^{-4}$. Similar experiments were performed with applied tones from 500 Hz to 400 kHz and while the results qualitatively agreed with theory, attenuation and distortion of the tones through various filters with uncharacterized parasitic capacitance and resistive loss at 4 K in our experimental setup prevented us from comparing quantitatively to the theory.\\
\\
Finally, we perform motion-echo experiments without a purposely applied tone, to sense and characterize intrinsic frequency noise in our setup. With $n_a=10$ (20 free-precession periods), $\tau_d=$ 4 $\mu$s, and coherent displacements $\Omega_x \tau_d /2 = 3.44(2)$, we find several peaks in the time scan corresponding to a single narrow-band noise spur at $\omega_n \simeq 2 \pi \times$ 260 kHz and with an amplitude $A_n = 2 \pi \times 2.4(2)$ kHz (see Fig. \ref{Fig:realnoise}a), which corresponds to a relative modulation depth $A_n/\omega_x \simeq 3\times 10^{-4}$. In this run, the electrode potentials are supplied from digital to analog converters (DACs), so the spurs are likely caused by cross-talk of digital circuitry in the DACs to the outputs. After switching the electrode potential sources to low-noise analog power supplies, we observed a nearly uniform noise floor without the spurs (see Fig. \ref{Fig:realnoise}b).
\section{Summary and conclusions}
We have theoretically described and experimentally demonstrated a number of features of coherently displaced harmonic oscillator (HO) number states, including displaced ground states. Coherent displacements are a universal concept that applies to all harmonic oscillators and, because they correspond to a classical, near-resonant force on the oscillator, they can often be implemented in a simple way in concrete experimental settings. Here, we characterized the responses of a single, harmonically trapped atomic ion to an electric field that oscillates close to resonance with the ion motion. The ion response was then characterized by coupling the HO to an internal two-level system of the ion. This was realized by driving a red sideband on an internal state transition that sensitively depends on the state of the HO and subsequently detecting state-dependent fluorescence from the atom. The ion fluorescence exhibits a non-linear response in $\bar{n}$ when driving the sideband for final states with average occupation number $\bar{n}> 1$. The resulting line shapes of internal-state population versus detuning exhibit a complicated but symmetric structure with steep and narrow side lobes to the central resonance feature which are accounted for by the theory.\\   
\\  
By applying a sequence of coherent displacements, alternating with free evolution of the state of motion, we obtained a frequency-filtered response of the HO. In this way, fluctuations of the HO frequency in a certain frequency band can be isolated and sensitively detected by the ion itself. We demonstrated the key features of this mechanism from the response to deliberately applied monochromatic modulations on the trap frequency, and then used the motion-echo sequence to detect and eliminate inherent technical HO frequency noise in our ion trap system.\\ 
\\
We anticipate that the basic concepts exhibited in this work can be transferred to many other HO systems and foresee various extensions and refinements of the current work. For example, it should be possible to find interesting modifications of the coherent displacement sequences by utilizing different displacement patterns in phase space or by displacing non-classical quantum states \cite{McCormick18,Wolf18}.\\
\\
%
%
\newpage
\section*{References}
\bibliographystyle{naturemag}
\bibliography{coherent_displacements}
\section*{Acknowledgements} 
We thank David Allcock, Shaun Burd and Daniel Slichter for helpful discussions and assistance with the experimental setup and Alejandra Collopy and Kevin Gilmore for useful comments on the manuscript. This work was supported by IARPA, ARO, ONR and the NIST quantum information program. K.C.M. acknowledges support by an ARO QuaCGR fellowship through grant W911NF-14-1-0079. J.K. acknowledges support by the Alexander von Humboldt Foundation.
\end{document}